\documentclass[twocolumn,prl,showpacs,preprintnumbers,
               superscriptaddress,amsmath,amssymb,floatfix]{revtex4}

\usepackage{color}
\usepackage{graphicx}% Include figure files
\usepackage{dcolumn}% Align table columns on decimal point
\usepackage{bm}% bold math
\usepackage{longtable}

\usepackage{epsfig,amssymb}

\newcommand{\br}{\bm{r}}

\begin{document}

%\begin{CJK*}{GBK}{}

\title{Neutron halo in deformed nuclei}

\author{Shan-Gui Zhou} %(\"{O}\"{U}\'{E}{\AE}¹\'{o})}
% \email{sgzhou@itp.ac.cn}
% \homepage{http://www.itp.ac.cn/~sgzhou}
 \affiliation{Key Laboratory of Frontiers in Theoretical Physics,
              Institute of Theoretical Physics, Chinese Academy of Sciences,
              Beijing 100190, China}
 \affiliation{Kavli Institute for Theoretical Physics China at the Chinese
              Academy of Sciences, Beijing 100190, China}
 \affiliation{Center of Theoretical Nuclear Physics, National Laboratory
              of Heavy Ion Accelerator, Lanzhou 730000, China}
\author{Jie Meng }%(\~{A}\"{I}½\"{U})}
% \email{mengj@pku.edu.cn}
 \affiliation{School of Physics, Peking University,
              Beijing 100871, China}
 \affiliation{Key Laboratory of Frontiers in Theoretical Physics,
              Institute of Theoretical Physics, Chinese Academy of Sciences,
              Beijing 100190, China}
 \affiliation{Kavli Institute for Theoretical Physics China at the Chinese
              Academy of Sciences, Beijing 100190, China}
 \affiliation{Center of Theoretical Nuclear Physics, National Laboratory
              of Heavy Ion Accelerator, Lanzhou 730000, China}
\author{P. Ring}
% \email{ring@physik.tu-muenchen.de}
 \affiliation{Physikdepartment, Technische Universit\"at M\"unchen,
              85748 Garching, Germany}
 \affiliation{School of Physics, Peking University,
              Beijing 100871, China}
 \affiliation{Kavli Institute for Theoretical Physics China at the Chinese
              Academy of Sciences, Beijing 100190, China}
\author{En-Guang Zhao}% (\~{O}\^{O}{\P}{\div}¹\~{a})}%
% \email{egzhao@itp.ac.cn}
 \affiliation{Key Laboratory of Frontiers in Theoretical Physics,
              Institute of Theoretical Physics, Chinese Academy of Sciences,
              Beijing 100190, China}
 \affiliation{Kavli Institute for Theoretical Physics China at the Chinese
              Academy of Sciences, Beijing 100190, China}
 \affiliation{Center of Theoretical Nuclear Physics, National Laboratory
              of Heavy Ion Accelerator, Lanzhou 730000, China}
 \affiliation{School of Physics, Peking University,
              Beijing 100871, China}

\date{\today}

\begin{abstract}
Halo phenomena in deformed nuclei are investigated within a deformed
relativistic Hartree Bogoliubov (DRHB) theory. These weakly bound
quantum systems present interesting examples for the study of the
interdependence between the deformation of the core and the particles
in the halo. Contributions of the halo, deformation effects, and
large spatial extensions of these systems are described in a fully
self-consistent way by the DRHB equations in a spherical Woods-Saxon
basis with the proper asymptotic behavior at large distance from the
nuclear center. Magnesium and neon isotopes are studied and detailed
results are presented for the deformed neutron-rich and weakly bound
nucleus $^{44}$Mg. The core of this nucleus is prolate, but the halo
has a slightly oblate shape. This indicates a decoupling of the halo
orbitals from the deformation of the core. The generic conditions
for the occurence of this decoupling effects are discussed.
\end{abstract}


%21.10.Gv    Nucleon distributions and halo features
%21.10.-k    Properties of nuclei; nuclear energy levels
%21.60.Jz    Nuclear Density Functional Theory and extensions
%            (includes Hartree¨CFock and random-phase approximations)
%21.60.-n    Nuclear structure models and methods

\maketitle

%\end{CJK*}

%\section{\label{sec:intro}Introduction}

The ``shape'' provides an intuitive understanding of spatial density
distributions in quantum many-body systems, such as
molecules~\cite{Sim.08}%\cite{Simons2008}
, atoms~\cite{Ceraulo1991}, atomic nuclei~\cite{Bohr1969}, or
mesons~\cite{Alexandrou2008}. Quadrupole deformations play an
important role in this context. The interplay between quadrupole
deformation and the weak binding can result in new phenomena, such
as ``quadrupole-bound'' anions~\cite{DBS.04x}.

Halo phenomena in nuclei are driving forces for the development of
the physics with radioactive ion beams. They are threshold
effects~\cite{JRF.04x} %\cite{Jensen2004}
and have been first observed in the weakly bound
system $^{11}$Li~\cite{THH.85bx}%\cite{Tanihata1985}
. Considering that most open shell nuclei are deformed, the
interplay between deformation and weak binding raises interesting
questions, such as whether or not there exist halos in deformed
nuclei and, if yes, what are their new features.

Calculations in a deformed single-particle model~\cite{MNA.97x} %\cite{Misu1997}
have shown that valence particles in specific orbitals with low
projection  of the angular momentum on the symmetry axis,
can give rise to halo structures in the limit of weak binding. The
deformation of the halo is in this case solely determined by the
intrinsic structure of the weakly bound orbitals. Indeed, halos in
deformed nuclei were investigated in several mean field calculations
in the past~\cite{Li1996,PXS.06x,Nakada2008}.
%\cite{Li1996, Pei2006, Nakada2008}.
However, in Ref.~\cite{Hamamoto2004}, it has been concluded that in
the neutron orbitals of an axially deformed Woods-Saxon potential
the lowest-$\ell$ component becomes dominant at large distances from
the origin and therefore all $\Omega^{\pi} = 1/2^+$ levels do not
contribute to deformation for binding energies close to zero. Such
arguments raise doubt about the existence of deformed halos. In
addition, a three-body model study~\cite{Nunes2005} suggests that it
is unlikely to find halos in deformed drip line nuclei because the
correlations between the nucleons and those due to static or dynamic
deformations of the core inhibit the formation of halos.

Therefore a model which provides an adequate description of halos in
deformed nuclei must include in a self-consistent way the continuum,
deformation effects, large spatial distributions, and the coupling
among all these features. In addition it should be free of adjustable
parameters that make predictions unreliable. Density functional
theory fulfills all these requirements. Spherical nuclei with halos
have been described in the past successfully in this way by the
solution of either the non-relativistic Hartree-Fock-Bogoliubov
(HFB)~\cite{Bulgac1980, Dobaczewski1984, Schunck2008} %
%\cite{Bulgac1980, Dobaczewski1984,Dobaczewski1996, Schunck2008}%
or the relativistic Hartree Bogoliubov (RHB) equations
~\cite{Meng1996,Poschl1997,Meng1998b} in coordinate ($r$) space.
However, for deformed nuclei the solution of HFB or RHB equations in
$r$ space is a numerically very demanding task. In the past
considerable effort has been made to develop mean field models
either in $r$ space or in a basis with an improved asymptotic
behavior at large
distances~\cite{THF.96x,Stoitsov2000,TOU.03x,Zhou2003a,Tajima2004,Stoitsov2008,Nakada2008}. %
%\cite{Terasaki1996,Stoitsov2000,Teran2003,Zhou2003a,Tajima2004,Stoitsov2008,Nakada2008}.%
In particular, an expansion in a Woods-Saxon (WS) basis was shown to
be fully equivalent to calculations in $r$ space~\cite{Zhou2003a}.
%\cite{Zhou2003a,Zhou2006, Zhou2008a}

In the present investigation, we therefore study halo phenomena in
deformed exotic nuclei within a DRHB model using a spherical WS
basis. The RHB equations for the nucleons
read~\cite{Kucharek1991,Meng1996}
\begin{equation}
 \hspace*{-0.4cm}
\left(
  \begin{array}{cc}
   h_D-\lambda&
   \Delta \\
  -\Delta^*
   & -h^*_D+\lambda\\
  \end{array}
 \right)
 \left(
  { U_{k}\atop V_{k} }
 \right)
 =
 E_{k}
  \left(
   { U_k\atop V_{k} }
  \right),
 \label{eq:RHB0}
\end{equation}
where $E_{k}$ is the quasiparticle energy, $\lambda$ the chemical
potential, and $h_D$ is the Dirac Hamiltonian
~\cite{Serot1986,Reinhard1989,Ring1996,Vretenar2005,Meng2006}
\begin{equation}
 h_D =
  \bm{\alpha} \cdot \bm{p} + V(\bm{r}) + \beta (M + S(\bm{r})).
  \label{eq:Dirac0}
\end{equation}
Neglecting here for simplicity spin and isospin degrees of freedom
the pairing potential reads,
\begin{equation}
 \Delta(\br_1,\br_2)= %\frac{1}{2}
 V^{pp}(\br_1,\br_2)\kappa(\br_1,\br_2),
\label{eq:pairing-potential}
\end{equation}
with a density dependent force of zero range in the particle-particle
channel
\begin{equation}
 V^{pp}(\br_1,\br_2)=  V_0\,\delta( \br_1 - \br_2 )
   \left(1-\frac{\rho(\br_1)}{\rho_\mathrm{sat}}\right)
   \frac{1}{2}(1-P^\sigma).
 \label{eq:pairing_force}
\end{equation}
and the pairing tensor $\kappa$ which is defined in the conventional
way.
%A  smooth cutoff in quasiparticle space depending on the parameters,
%$E_\mathrm{cut}$ and $\Gamma_\mathrm{cut}$, similar to the soft
%cutoff proposed in Ref.~\cite{Bonche1985} , is adopted:
%\begin{equation}
% f_k =\frac{1}{2} \left( 1 - \frac{E_k-E_\mathrm{cut}}
% {\sqrt{(E_k-E_\mathrm{cut})^2+(\Gamma_\mathrm{cut})^2}} \right).
% \label{eq:smooth}
%\end{equation}

For axially deformed nuclei with spatial reflection symmetry, we
represent the potentials and densities in terms of the Legendre
polynomials,
\begin{equation}
 f(\bm{r})   = \sum_\lambda f_\lambda({r}) P_\lambda(\cos\theta),\
 \lambda = 0,2,4,\cdots.
 \label{eq:expansion}
\end{equation}
For fixed quantum numbers $\Omega^\pi$ the deformed quasiparticle
wave functions $U_k(\br)$ and $V_k(\br)$ in Eq.~(\ref{eq:RHB0}) are
expanded in a spherical WS basis (for details see
Ref.~\cite{Zhou2003a}).

\begin{figure}
\includegraphics[width=8cm]{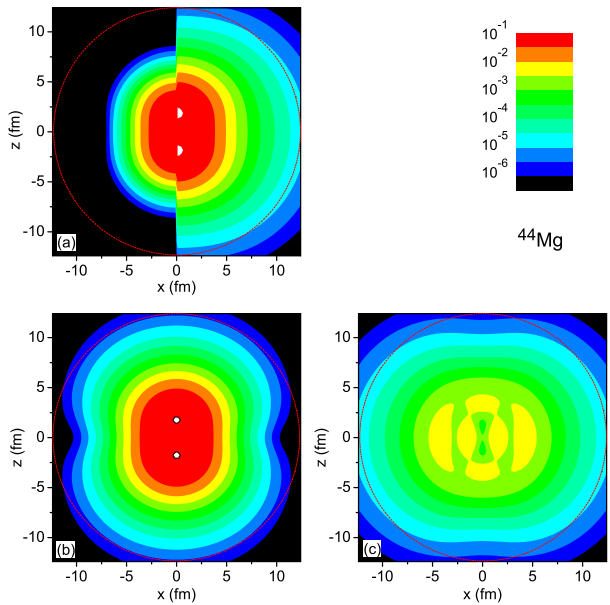}
\caption{\label{fig1}%
(Color online) Density distributions of $^{44}$Mg with the $z$-axis
as symmetry axis: (a) the proton density (for $x<0$) and the neutron
density (for $x>0$), (b) the density of the neutron core, and (c)
the density of the neutron halo. In each plot, a dotted circle is
drawn for guiding the eye.}
\end{figure}

The calculations are based on the density functional
NL3~\cite{Lalazissis1997} and the $pp$
interaction~(\ref{eq:pairing_force})
% and(\ref{eq:smooth})%
with the parameters $\rho_\mathrm{sat} =$ 0.152~fm$^{-3}$, $V_0 =
380$~MeV$\cdot$fm$^3$, and a cut-off energy $E^\mathrm{q.p.}_\mathrm{cut} =
60$~MeV in the quasi-particle space.
%and $\Gamma_\mathrm{cut} = 5.65$ MeV
These parameters reproduce the proton pairing energy of the
spherical nucleus $^{20}$Mg obtained from a spherical RHB
calculation with the Gogny force D1S. A spherical box of the size
$R_\mathrm{max} = 20$ fm and the mesh size $\Delta r = 0.1$ fm are
used for generating the spherical Dirac WS basis of
Ref.~\cite{Zhou2003a} which consists of states with $j< \frac{21}{2}
\hbar$. An energy cutoff $E^+_\mathrm{cut}$ = 100 MeV is applied to
truncate the positive energy states in the WS basis and the number
of negative energy states in the Dirac sea is taken to be the same
as that of positive energy states in each ($\ell,j$)-block.

%In the present study of Mg isotopes, the last nucleus within the
%neutron drip-line is $^{46}$Mg, an almost spherical nucleus.
In the present study of Mg isotopes, the last nucleus within the
neutron drip-line is $^{46}$Mg. Of course, it is difficult to predict
the position of the drip-line precisely for nuclei so far from the
experimentally known area and therefore the results discussed in the
following have to be taken as generic results. In this study
$^{46}$Mg is an almost spherical nucleus. The neighboring nucleus
$^{44}$Mg is well deformed ($\beta_2 = 0.32$) and weakly bound with
the two-neutron
separation energy $S_{2n} = 0.44$ MeV. Therefore this nucleus is
taken here as an example for a detailed investigation. The density
distributions of all protons and all neutrons in $^{44}$Mg are shown
in Fig.~\ref{fig1}a. Due to the large neutron excess, the neutron
density not only extends much farther in space but it also shows a
halo structure. The neutron density is decomposed into the
contribution of the core in Fig.~\ref{fig1}b and that of the halo in
Fig.~\ref{fig1}c. Details of this decomposition are given further
down. We find that the core of $^{44}$Mg is prolate, and that the
halo has a slightly oblate deformation. This indicates the decoupling
between the deformations of core and halo.

\begin{figure}
\includegraphics[width=8.0cm]{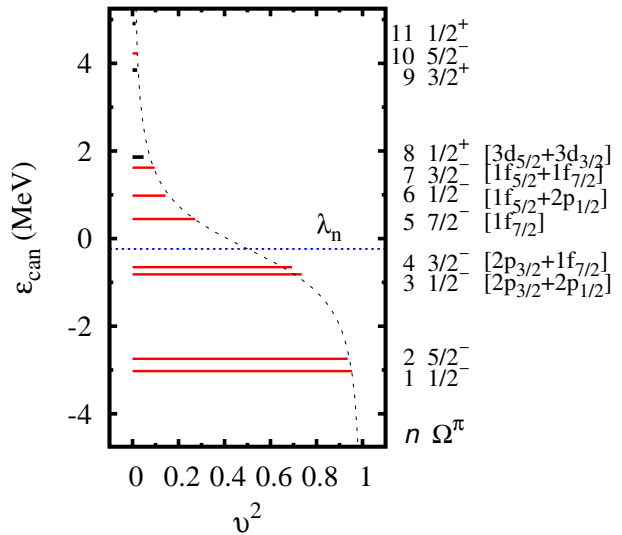}%
\caption{\label{fig2} (Color online) Single neutron levels with the
quantum numbers $\Omega^\pi$ around the chemical potential (dotted
line) in the canonical basis for $^{44}$Mg as a function of the
occupation probability $v^2$. The order $n$, $\Omega^\pi$, and the
main WS components for orbitals close to the threshold are also
given. The dashed line corresponds to the BCS-formula with an average
pairing gap.}
\end{figure}

Weakly bound orbitals or those embedded in the continuum play a
crucial role in the formation of a nuclear halo. For an intuitive
understanding of the single particle structure we keep in mind that
HB-functions can be represented by BCS-functions in the canonical
basis and show in Fig.~\ref{fig2} the corresponding single neutron
spectrum. As discussed in Ref.~\cite{RS.80} the single particle
energies in the canonical basis $\varepsilon_k=\langle
k|h_D|k\rangle$ shown in Fig.~\ref{fig2} are expectation values of
the Dirac Hamiltonian~(\ref{eq:Dirac0}) for the eigenstates
$|k\rangle$ of the single particle density matrix $\hat\rho$ with the
eigenvalues $v^2_k$. The spectrum of $\hat\rho$ has a discrete part
with $v^2_k>0$ and a continuous part with $v^2_k=0$. Obviously only
the first part contributes to the HB-wave function and only this part
is plotted in Fig.~\ref{fig2}. This part of the spectrum
$\varepsilon_k$ is discrete even for the levels in the continuum. Of
course, this is only possible because the wave functions $|k\rangle$
are not eigenfunctions of the Hamiltonian. As long as the chemical
potential $\lambda_\mathrm{n}$ is negative, the corresponding density
$\rho(\br)$ is localized~\cite{Dobaczewski1984} and the particles
occupying the levels in the continuum are bound.

The orbitals in Fig.~\ref{fig2} are labeled by the conserved quantum
numbers $\Omega$ and $\pi$. The character $n$ numbers the different
orbitals appearing in this figure according to their energies. The
neutron Fermi energy lies within the $pf$ shell and most of the single
particle levels have negative parities. Since the chemical potential
$\lambda_\mathrm{n} = -230$ keV is relatively small, orbitals above
the threshold have noticeable occupation due to pairing correlations.
For example, the occupation probabilities of the 5th ($\Omega^\pi =
7/2^-$) and the 6th ($\Omega^\pi = 1/2^-$) orbitals are 27.2\% and
14.3\%.

\begin{figure}
\includegraphics[width=7cm]{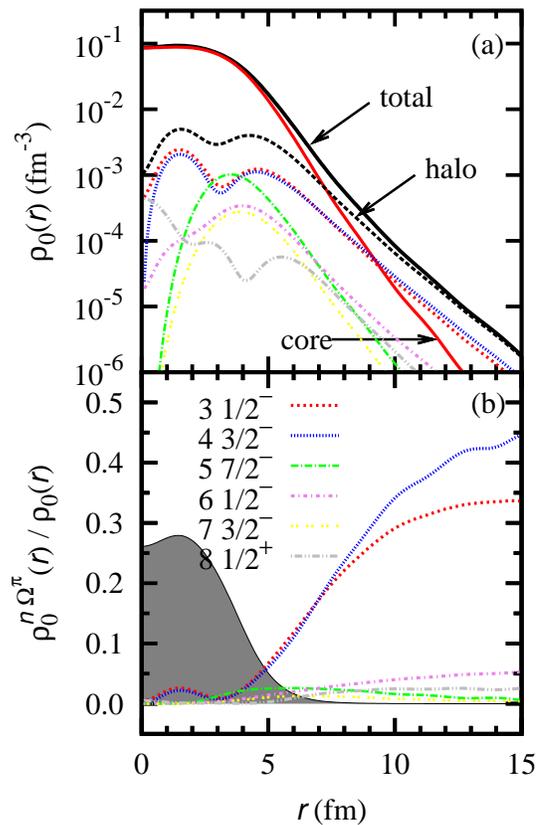}
\caption{\label{fig3} (Color online) Neutron density distributions
for $^{44}$Mg. (a) The total density and its decomposition into core
and halo. Contributions from several neutron orbitals around the
Fermi level are also given. (b) Relative contributions of these
neutron orbitals to the total neutron density, which is indicated in
arbitrary units by the shaded area.}
\end{figure}

As we see in Fig.~\ref{fig2} there is a considerable gap between the
two levels with the numbers $n=2$ and $n=3$. The levels with
$\varepsilon_\mathrm{can} < -2.5$ MeV contribute to the ``core'', and
the other remaining weakly bound and continuum orbitals with
$\varepsilon_\mathrm{can} > -1$ MeV naturally form the ``halo''.
Therefore we decompose the neutron density $\rho^n(\br)$ into two
parts, one part coming from the orbitals with canonical single
particle energies $\varepsilon_\mathrm{can} < -2.5$ MeV (called
``core'') and the other from the remaining weakly bound and continuum
orbitals (called ``halo''). The spherical components of these
densities [i.e. the contribution of $\lambda=0$
in~Eq.~(\ref{eq:expansion})] are plotted together with that of the
total neutron density in Fig.~\ref{fig3}a. It is seen that the tail
part of the neutron density originates mainly from the orbitals with
$\varepsilon_\mathrm{can} > -1$ MeV. The average number of neutrons
which are weakly bound or in the continuum is around 4.34. On the
average, 2.92 of these neutrons are in the weakly bound orbits 3 and
4 and the others in the continuum. The rms radii of the core and the
halo are 3.72 fm and 5.86 fm, respectively. A further decomposition
shows that the two weakly bound orbitals, i.e., the 3rd ($\Omega^\pi
= 1/2^-$) and the 4th ($\Omega^\pi = 3/2^-$), contribute mostly to
the halo. This is more clearly seen in Fig.~\ref{fig3}b where we
represent the relative contributions of weakly bound and continuum
orbitals to the total neutron density. The two continuum orbitals,
i.e., the 6th ($\Omega^\pi = 1/2^-$) and the 8th ($\Omega^\pi =
1/2^+$) also contribute to the tail.

\begin{figure}[b]
\includegraphics[width=7cm]{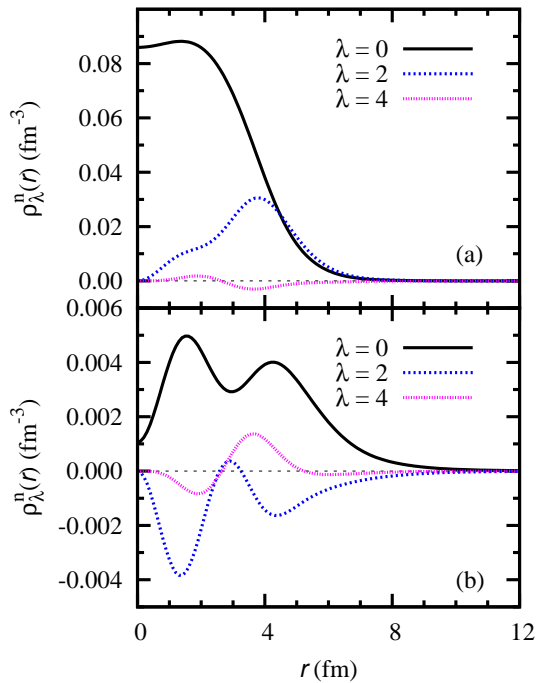}
\caption{\label{fig4} (Color online) Decomposition of the neutron
density of $^{44}$Mg into spherical ($\lambda=0$), quadrupole
($\lambda=2$), and hexadecapole ($\lambda=4$) components for (a) the
core and (b) the halo. }
\end{figure}

If we decompose the deformed wave functions of the two weakly bound
orbitals, i.e. the 3rd ($\Omega^\pi = 1/2^-$) and the 4th
($\Omega^\pi = 3/2^-$), in the spherical WS basis it turns out that
in both cases the major part comes from $p$ waves as indicated on the
right hand side of Fig.~\ref{fig2}. The $p$ wave components for the
3rd and the 4th orbitals are 66\% and 80\% respectively. Having in mind
that the occupation probabilities of these two orbitals are 0.736 and
0.693 and each orbital has degeneracy 2, there are about 2 neutrons
in weakly bound $p$ states. The low centrifugal barrier for $p$ waves
gives rise to the formation of the halo. Having a small $p$ wave
component, the 6th orbital ($\Omega^\pi = 1/2^-$) contributes less to
the halo though it is in the continuum and the occupation probability
is rather large. The contribution of the 8th orbital ($\Omega^\pi =
1/2^+$) to the tail of the density is even smaller because its main
components are of $d$ waves. The large centrifugal barrier of $f$ states
hinders strongly the spatial extension of the wave functions of the
other two continuum orbitals, i.e., the 5th ($\Omega^\pi = 7/2^-$)
and the 7th ($\Omega^\pi = 3/2^-$).

\begin{figure}
\includegraphics[width=8cm]{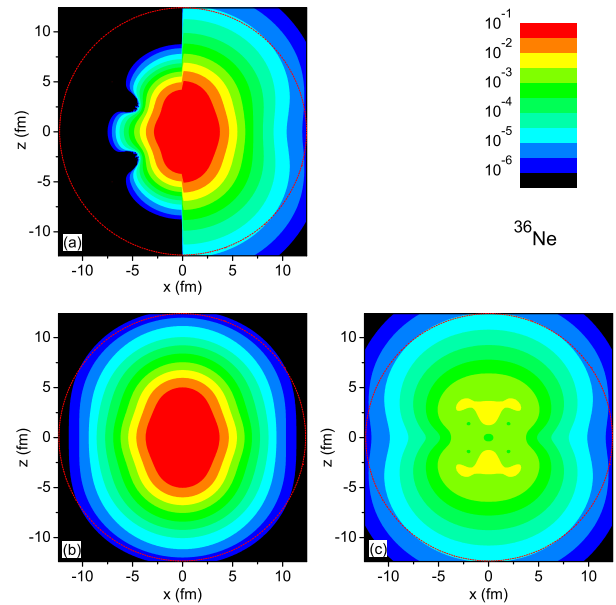}
\caption{\label{fig5}%
(Color online) Density distributions of $^{36}$Ne. Details are given
in Fig. 1}
\end{figure}

In Fig.~\ref{fig4} the densities of the core and the halo are decomposed
into spherical, quadrupole, and hexadecapole components.
As is seen in Fig.~\ref{fig4}a, the quadrupole component of the core
is positive, thus being consistent with the prolate shape of
$^{44}$Mg. However, for the halo, the quadrupole component has a
negative sign, which means that the halo has an oblate deformation.
The quadrupole moments of the neutron core and the halo are 160 and
$-$27 fm$^2$, respectively. This explains the decoupling between the
quadrupole deformations of the core and the halo as we have seen it
in Figs.~\ref{fig1}b and~\ref{fig1}c. There is also a noticeable
hexadecapole component in the density distribution of the halo.

The slightly oblate shape of the halo originates from the intrinsic
structure of the weakly bound and continuum orbitals. As is shown in
Fig.~\ref{fig2}, the main WS components of the two weakly bound
orbitals, the 3rd ($\Omega^\pi = 1/2^-$) and the 4th ($\Omega^\pi =
3/2^-$), are $p$ states. We know that the
angular distribution of $|Y_{10}(\theta,\phi)|^2
\propto \cos^2\theta$ with a projection of the orbital angular
momentum on the symmetry axis $\Lambda=0$ is prolate and that of
$|Y_{1\pm1}(\theta,\phi)|^2 \propto \sin^2\theta$ with
$\Lambda=1$ is oblate. It turns out that in the 3rd
($\Omega^\pi = 1/2^-$) orbital, both $\Lambda = 0$ and $\Lambda = 1$
components contribute and the latter dominates. Therefore this orbital
has a slightly oblate shape. For the 4th ($\Omega^\pi = 3/2^-$) state,
there is only the $\Lambda=1$ component from the $p_{3/2}$ wave,
an oblate shape is also expected.

\begin{figure}
\includegraphics[width=8cm]{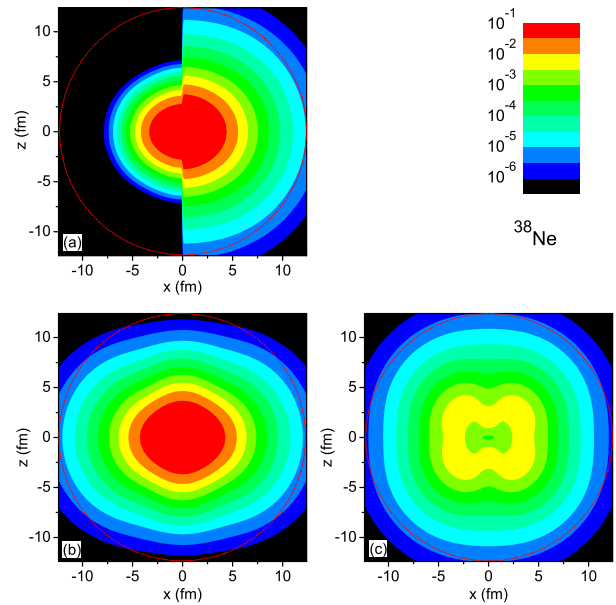}
\caption{\label{fig6}%
(Color online) Density distributions of $^{38}$Ne. Details are given
in Fig. 1}
\end{figure}

In order to show that these results depend crucially on the single
particle structure in the neighborhood of the Fermi surface we also
investigate weakly-bound nuclei in the neighboring chain of
Ne-isotopes. In Fig.~\ref{fig5}a the density distributions of all
protons and all neutrons in the prolate deformed nucleus $^{36}$Ne
are shown ($\beta_2 = 0.52$). Again, as in $^{44}$Mg, due to the
large neutron excess, the neutron density not only extends much
farther in space but it also shows a halo structure. The neutron
density is decomposed into the contribution of the core in
Fig.~\ref{fig5}b and that of the halo in Fig.~\ref{fig5}c. In
contrary to the nucleus $^{44}$Mg, we observe now a prolate halo,
because the essential level of the halo has a large contribution from
the prolate $\Lambda = 0$ ($p$ wave) component. In Fig.~\ref{fig6} we
show similar density distributions for the oblate deformed nucleus
$^{38}$Ne ($\beta_2 = -0.24$) which is the last nucleus within the
neutron drip line in the present calculation. In this case the Fermi
level is again within the $pf$ shell. But the levels dominated by $p$
waves are either less occupied or not so weakly bound and therefore
we do not find a pronounced halo. From these examples it is clear
that the existence and the deformation of a possible neutron halo
depends essentially on the quantum numbers of the main components of
the single particle orbits in the vicinity of the Fermi surface:
$s$ levels with $\Lambda=0$ produce spherical halos, $p$ levels
with $\Lambda=0$ prolate and $p$ levels with $\Lambda=1$ oblate
halos~\cite{MNA.97x}.

In summary, the very neutron-rich deformed nucleus $^{44}$Mg is
investigated within deformed relativistic Hartree Bogoliubov theory
in the continuum. In contrast to several
expectations~\cite{Hamamoto2004,Nunes2005} a pronounced deformed
neutron halo is found. It is formed by several orbitals close to the
threshold (either weakly bound or in the continuum). They have large
components of low $\ell$-values and feel therefore only a small
centrifugal barrier. Although $^{44}$Mg and its core are well
deformed and prolate, the deformation of the halo is slightly
oblate. This implies a decoupling between the deformations of core
and halo. This mechanism is investigated by studying the details of
the neutron densities for core and halo, the single particle levels
in the canonical basis, and the decomposition of the halo orbitals.
We also studied the weakly-bound nuclei in Ne isotopes and discussed
the conditions for the occurence of a halo and its shape. It is
shown that the existence and the deformation of a possible neutron
halo depends essentially on the quantum numbers of the main
components of the single particle orbits in the vicinity of the
Fermi surface.

%All these studies show that irrespective of the shape of the core,
%the deformation of the halo originates from the intrinsic properties
%of those orbitals which mostly contribute to the formation of the
%halo.

Finally we note that besides the ``quadrupole-bound''
molecule~\cite{DBS.04x} %\cite{Desfrancois2004}
and the nuclear halo in deformed nuclei, similar coupling effects
between the deformation and the weakly bound part of the system could
also exist in other quantum many-body systems, such as Rydberg atoms
in which the electron(s) can be extremely weakly bound and where the
quadrupole moment is sizable~\cite{Ceraulo1991}.

This work has been supported in part by Natural Science Foundation
of China (10775004, 10705014, 10875157, and 10979066), Major State Basic
Research Development Program of China (2007CB815000), Knowledge
Innovation Project of Chinese Academy of Sciences (KJCX3-SYW-N02 and
KJCX2-YW-N32), by the Bundesministerium f\"{u}r Bildung und
Forschung (BMBF), Germany, under Project 06 MT 246, and by the DFG
cluster of excellence \textquotedblleft Origin and Structure of the
Universe\textquotedblright\ (www.universe-cluster.de). The
computation was supported by Supercomputing Center, CNIC of CAS. One
of the authors (P.R.) would like to express his gratitude to J. Meng
for the kind hospitality extended to him at the Peking University.
Helpful discussions with N. V. Giai, Z. Y. Ma, N. Sandulescu, J.
Terasaki, D. Vretenar, S. J. Wang, and S. Yamaji are gratefully
acknowledged.

%\bibliographystyle{apsrev}
%\bibliography{sgzhou}
%\bibliography{../../../information/refs/JabRef/sgzhou}

%\bibliographystyle{c:/B/a00/prsty}

%\bibliography{c:/B/a00/refring}

\end{document}